\documentclass[12pt]{iopart}

\expandafter\let\csname equation*\endcsname\relax
\expandafter\let\csname endequation*\endcsname\relax
\usepackage{iopams}  
\usepackage{soul}
\usepackage{amsmath,amssymb}
\usepackage{lineno}
\usepackage{enumitem}
\usepackage{graphicx}
\begin{document}

%% ------------------------------------------------------------------------ %%
%  Title
%
% (A title should be specific, informative, and brief. Use
% abbreviations only if they are defined in the abstract. Titles that
% start with general keywords then specific terms are optimized in
% searches)
%
%% ------------------------------------------------------------------------ %%

\title[Tropical teleconnections and midlatitudes predictability]{Impact of tropical teleconnections on the long-range predictability of the atmosphere at midlatitudes: A reduced-order multi-scale model perspective}

%% ------------------------------------------------------------------------ %%
%
%  AUTHORS AND AFFILIATIONS
%
%% ------------------------------------------------------------------------ %%

% Authors are individuals who have significantly contributed to the
% research and preparation of the article. Group authors are allowed, if
% each author in the group is separately identified in an appendix.)

% List authors by first name or initial followed by last name and
% separated by commas. Use \affil{} to number affiliations, and
% \thanks{} for author notes.
% Additional author notes should be indicated with \thanks{} (for
% example, for current addresses).

% Example: \authors{A. B. Author\affil{1}\thanks{Current address, Antartica}, B. C. Author\affil{2,3}, and D. E.
% Author\affil{3,4}\thanks{Also funded by Monsanto.}}

\author{St\'ephane Vannitsem}

\address{Royal Meteorological Institute of Belgium, Avenue Circulaire, 3, 1180 Brussels, Belgium}

\eads{Stephane.Vannitsem@meteo.be}

%% Corresponding Author:
% Corresponding author mailing address and e-mail address:

% (include name and email addresses of the corresponding author.  More
% than one corresponding author is allowed in this LaTeX file and for
% publication; but only one corresponding author is allowed in our
% editorial system.)

% Example: \correspondingauthor{First and Last Name}{email@address.edu}

%\correspondingauthor{St\'ephane Vannitsem}{Stephane.Vannitsem@meteo.be}

%% Keypoints, final entry on title page.

%  List up to three key points (at least one is required)
%  Key Points summarize the main points and conclusions of the article
%  Each must be 100 characters or less with no special characters or punctuation and must be complete sentences

\bigskip
\bigskip

\begin{indented}
    \item May 20, 2023
\end{indented}

\bigskip
\bigskip

\begin{abstract}
Teleconnections between the tropical and the extratropical climates are often considered as a potential source of long-term predictability at seasonal to decadal time scales in the extratropics. This claim is taken up in the present work by investigating the predictability of a coupled ocean-atmosphere extratropical model under a one-way forcing generated by a tropical model. Both models display a chaotic dynamics, and the dominant variable of the extratropical model displays a high correlation with the tropical forcing in the reference simulation, inducing a low-frequency variability signal in the extratropics. 

Numerical experiments emulating the presence of initial condition errors in the tropical model are conducted to clarify their impact on the predictability in the extratropics. It is shown that: (i) The correlation skill of the dominant observable affected by the forcing is considerably degraded at interannual time scales due to the presence of initial condition errors in the tropics, considerably limiting the potential of teleconnections; (ii) averaging of an ensemble of forecasts -- with a small number of members -- may substantially improve the quality of the forecasts; and (iii) temporal averaging may also improve the quality of the forecasts (at the expense of being able to forecast extreme events), in particular when the forcing affects weakly the observable under interest.
	
\end{abstract}


\newpage

\section{Introduction}

The tropical regions display key low-frequency variability ranging from months with the Madden-Julian Oscillation to  annual and decadal time scales for the El-Niño-Southern Oscillation (ENSO). These processes display teleconnections with large regions in the extratropics as discussed in \cite{Philander1990, Alexanderetal2002, Hoerling2002, Stan2017, Yeh2018}. These are often expressed in terms of correlations between the tropical and extratropical regions, suggesting a potential influence on one another. Note however than correlation does not mean causation as already stressed in many recent works on causality analyses \cite{Mosedale2006, Palus2018, Runge2019, Vannitsem2022b}. In \cite{Vannitsem2022b}, a new approach of causality detection was used to clarify the transfer of information among different climate indices and in particular ENSO on monthly to decadal time scales. It was in particular shown that ENSO has considerable influence on the Pacific north Atlantic pattern, on the Pacific decadal oscillation, on the tropical north Atlantic region and slightly on the Arctic oscillation. The mechanism often put forward for the teleconnections with the midlatitudes is the atmospheric bridge described in \cite{Alexanderetal2002}. 

As these tropical processes have typical time scales of evolution much longer than the ones of the extratropical dynamics, it is natural to expect that these low-frequency variability signals would emerge in the evolution of the extratropical atmosphere, and could be exploited in our seasonal and decadal forecasts of the midlatitude dynamics \cite{KumarandHoerling1995, Nidheeshetal2017}, although the way these frequencies will emerge is a complicate matter as illustrated in \cite{Vannitsem2022}. But the processes developing in the tropical regions are not simple periodic signals that could be easily exploited as for instance the annual cycle. They display an erratic behavior reminiscent of chaotic or stochastically-forced dynamics \cite{Timmermannetal2003}. This implies that these processes display the property of sensitivity to initial conditions that will affect any forecast of the tropical evolution as initial state errors are always present in the model forecasts, and these uncertainties will, in turn, affect the evolution of the signal at midlatitudes. 

The starting point of the present work is the claim that any forecast will hardly be able to exploit the full potential of teleconnections. This claim is demonstrated by investigating the impact of errors in the initial conditions of a tropical forcing on an extratropical model. This system has been recently built to investigate the emergence of pullback attractors in a midlatitude reduced-order coupled ocean-atmosphere model forced by a tropical low-order model \cite{Vannitsem2022}. The midlatitude model is the so-called VDDG model (for Vannitsem-Demaeyer-DeCruz-Ghil) describing the interaction between a 2-layer quasi-geostrophic atmosphere with a 1.5-layer quasi-geostrophic ocean \cite{Vannitsem2015}, while the tropical low-order model is a box model developed by \cite{Jin1996, Jin1997, Timmermannetal2003}. A one-way interaction is imposed from the tropical model to the extratropical one. The solutions of both models that will be used in the present work display chaotic dynamics, and are therefore sensitive to the initial conditions. The experiments that will be performed will be to introduce random initial state errors of different amplitudes in the tropical model to clarify the ability of the extratropical model to produce long-term skillful forecasts at midlatitudes. 

In Section 2, the coupled tropical-extratropical model setup is described, together with the parameter used in the current experiment and the reference solutions around which the predictability experiments will be performed. Section 3 is devoted to the description of the experimental setup and the analysis of the different experiments. The impact of ensemble averaging and temporal averaging are also explored in this section. The main conclusions are drawn in Section 4.

\section{Governing equations for the tropical--extratropical model}
\label{sec:equate}

\subsection{The ENSO module}
\label{ssec:ENSO}

The ENSO model was developed in a series of papers by F.-F. Jin, A. Timmermann and colleagues \cite{Jin1996, Jin1997, AnJin2004, Timmermannetal2003, Robertsetal2016}. They  modeled the dynamics of the ocean's upper layer in the tropical Pacific using a low number of variables. Their ENSO model describes the dynamics of the temperature in the eastern and western tropical Pacific basins, completed by an equation for the evolution of the thermocline depth. The model represents the horizontal discharge-recharge mechanisms at play in the tropical Pacific through the heat exchanges between the tropical and subtropical waters, subject to surface wind stress and upwelling of subsurface cold water in the eastern part of the domain \cite{Jin1997}. \cite{Robertsetal2016} developed a nondimensional model version in which the time, for instance, is normalized by a typical time scale of tropical wave propagation of roughly 3.5 months. The latter ENSO model version is coupled here with an extratropical module whose time is nondimentionalized by the Coriolis parameter $f_0$, and the \cite{Robertsetal2016} equations are therefore slightly modified in order to incorporate this change of reference time scale:      

\begin{subequations} \label{eq:ENSO}
\begin{align}
\frac{dx}{dt} & =  \rho \delta (x^2-a x) + s x (x+y+c-c \tanh(x+z)), \label{ENSO1} \\
\frac{dy}{dt} & =  -\rho \delta (a y+x^2), \label{ENSO2} \\  
\frac{dz}{dt} & =  \delta (k-z-\frac{x}{2}) \label{ENSO3}.
\end{align}  
\end{subequations}

Here $x, y, z$ and $t$ are dimensionless, $x$ is the temperature difference between the eastern and western basins of the tropical Pacific, $y$ the western basin's temperature anomaly with respect to a reference value, and $z$ the western basin's thermocline depth anomaly.

The dimensionless parameters are defined as follows

\begin{subequations} \label{eq:dimens}
\begin{align}
a & = \frac{ \alpha b L}{\epsilon h^{\*} \beta}, \qquad  \rho = \frac{\epsilon h^{\*} \beta}{r b L}, 
\qquad   \delta = \frac{r}{f_0}, \label{dimens-1} \\
c & = \frac{C}{S_0}, \qquad k = \frac{K}{S_0},  \qquad  s = \frac{\zeta h^{\*} \beta}{b L f_0}, \label{dimens-2} 
\end{align}  
\end{subequations}

with $f_0$ the Coriolis parameter used to nondimensionalize the time, and the other parameters as defined in \cite[Table~1]{Robertsetal2016}. For the applications presented in this paper, one set of parameter values is considered leading to a chaotic solution, see Table \ref{tab:param}.

\begin{table}[ht!]
\centering
\caption{Dimensionless parameter values of the ENSO module}
%% \normalsize  % reset to normal font size
\setlength\tabcolsep{6 pt} % increase column separation
\begin{tabular}{l} % no vertical lines
\hline
\\
  $a=7.658609809$  \\
 $\rho =0.29016$       \\
 $\delta=0.0002803$     \\
$c=2.3952$   \\
 $k=0.4032$ \\
 $s=0.001069075$ \\
\hline
\end{tabular} \label{tab:param}
\end{table}

The ENSO module is taken here to be unaffected by the midlatitude module, and is thus the driving system in the coupled tropical-extratropical model. Equations~\eqref{eq:ENSO} can thus be integrated independently of the rest of the model, and this is done using a fourth-order Runge-Kutta scheme with a time step $\Delta t = 0.1346$~hours $=0.05$ nondimensional time units. The variable $x + y$ corresponds in the model \eqref{eq:ENSO} to the sea surface temperatures in the eastern tropical Pacific that are commonly associated with the Ni\~no-3 index, that will be referred later as $T_{ENSO}$. A long chaotic solution was started from the initial state $(x = -2.8439, y = -0.62, z = 1.480)$ that will be used in the following as the long reference trajectory of the tropical dynamics.

\subsection{The VDDG extratropical model} \label{ssec:VDDG}

The coupled ocean--atmosphere model used herein for the midlatitudes was developed by \cite{Vannitsemetal2015} and it is called hereafter the VDDG model. Different versions of this model have been used to study low-frequency variabiliy (LFV) within the coupled ocean--atmosphere system \cite{DeCruzetal2016,Vannitsem2017}, to investigate the properties of the Lyapunov exponents in such a system \cite{VannitsemLucarini2016, DeCruzetal2018}, to build the stochastic parametrization of subgrid-scale forcing \cite{Demaeyer2017, Demaeyer2018}, and to develop data assimilation schemes in coupled models \cite{Pennyetal2019, Tondeuretal2020, Carrassietal2022}. 

The atmospheric module is based on the vorticity equations of a two-layer quasi-geostrophic flow defined on a beta-plane \cite{Gill.1982, Pedlosky.1987}. The equations in pressure coordinates are:

\begin{subequations} \label{eq:VDDG}
\begin{align}
& \frac{\partial}{\partial t} \left( \nabla^2 \psi^1_a \right) + J(\psi^1_a, \nabla^2 \psi^1_a) + \beta \frac{\partial \psi^1_a}{\partial x}
 = - k'_d \nabla^2 (\psi^1-\psi^3) + \frac{f_0}{\Delta p} \omega, \label{VDDG-1} \\
& \frac{\partial}{\partial t} \left( \nabla^2 \psi^3_a \right) + J(\psi^3_a, \nabla^2 \psi^3_a) + \beta \frac{\partial \psi^3_a}{\partial x}
 =  + k'_d \nabla^2 (\psi^1_a-\psi^3_a) - \frac{f_0}{\Delta p}  \omega  
- k_d \nabla^2 (\psi^3_a-\psi_o). \label{VDDG-2} 
\end{align}  
\end{subequations}

Here $\psi^1_a$ and $\psi^3_a$ are the streamfunction fields at 250~hPa and 750~hPa, respectively, while $ \omega =dp/dt$ is the vertical velocity, $f_0$ the Coriolis parameter at $\phi_0 =$~45 $^{\circ}$ latitude, and $\beta = df/dy$ at $\phi_0$. The coefficients $k_d$ and $k'_d$ multiply the surface friction term and the internal friction between the layers, respectively. 

An additional term has been introduced in Eq.~\eqref{VDDG-2} in order to account for the presence of a surface boundary velocity $\psi_o$ of the oceanic flow; see Eq.~\eqref{eq:ocean} below. This term corresponds to the Ekman pumping on a moving surface and is the mechanical contribution of the interaction between the ocean and the atmosphere.

The ocean module is described by the reduced-gravity, quasi-geostrophic shallow-water model \cite{Gill.1982, Pedlosky.1987}. The forcing is provided by the wind generated by the atmospheric module above. The governing equation is:

\begin{equation} \label{eq:ocean}
\frac{\partial}{\partial t} \left( \nabla^2 \psi_o - \frac{\psi_o}{L_R^2} \right) + J(\psi_o, \nabla^2 \psi_o) + \beta \frac{\partial \psi_o}{\partial x}
= -r \nabla^2 \psi_o + \frac{{\mathrm{curl}}_z \vec{\tau}}{\rho H},
\end{equation}

where $\psi_o$ is the streamfunction, $\rho$ the density of water, $H$ the depth of the fluid layer, $L_R$ the reduced Rossby deformation radius, $r$ a Rayleigh friction coefficient at the bottom of the fluid layer, and curl$_z \vec{\tau}$ is the vertical component of the wind stress curl.

The wind stress in the VDDG model is given by $(\tau_x, \tau_y)=C (u-U,v-V)$, where $(u = -\partial \psi^3_{\rm a}/\partial y, v = \partial \psi^3_{\rm a}/\partial x)$ are the horizontal components of the geostrophic wind, and $(U, V)$ the components of the geostrophic currents in the ocean. One thus gets

\begin{equation}
	\mathrm{curl}_z \tau = C \nabla^2 (\psi^3_{\rm a}-\psi_{\rm o}). 
	\label{eq:stress}
\end{equation}

and the wind stress is proportional to the relative velocity between the flow in the ocean's upper layer and the wind in the lower atmospheric layer. The drag coefficient $d = C/(\rho_0 H)$ gives the strength of the mechanical coupling between the ocean and the atmosphere and it was a key bifurcation parameter in \cite{Vannitsemetal2015}.

The dynamic equations \eqref{eq:VDDG} and \eqref{eq:ocean} are supplemented by temperature equations for the two subsystems. For the atmosphere, 

\begin{equation} \label{eq:temp_a}
\gamma_a \left( \frac{\partial T_a}{\partial t} + J(\psi_a, T_a) -\sigma \omega \frac{p}{R}\right) = -\lambda (T_a-T_o) + E_{a,R},
\end{equation}

with

\begin{equation} \label{eq:rad_a}
E_{a,R} = \epsilon_a \sigma_B T_o^4 - 2 \epsilon_a \sigma_B T_a^4 + R_a.
\end{equation}

Here $R$ is the gas constant, $\epsilon_a$ the emissivity of the atmosphere, $\sigma_B$ the Stefan-Boltzman constant, $R_a$ the shortwave radiation at the top of the atmosphere, $\omega$ the vertical velocity in pressure coordinates, and $\sigma =  -(R/p) (\partial T_a/\partial p - 1/(\rho_a c_p))$ is the static stability, with $p$ the pressure, $\rho_a$ the air density, $c_p$ the specific heat at constant pressure, and $\sigma$ here is taken to be a constant. Note also that, thanks to the hydrostatic relation in pressure coordinates and the
ideal gas relation $p=\rho_a R T_a$, the atmospheric temperature can be written as
$T_a = - (p/R) f_0 (\partial \psi_a/\partial p)$. %% where $f_0$ is the Coriolis parameter at 45\textdegree  ~latitude. 

\begin{table}[ht!]
	\centering
	\caption{List of parameters of the extratropical VDDG module}
	%% \begin{tabular}{l r|l r}
	\setlength\tabcolsep{6 pt} % increase column separation
	\begin{tabular}{llll} % no vertical lines
		\hline 
		Parameter (unit) & Value & Parameter (unit) & Value \\
		\hline
		$L_y = \pi L$  (km)         & $5.0 \times 10^3$        & $\gamma_{\rm{o}}$ (J\,m$^{-2}$\, K$^{-1}$) & $4 \times 10^6 \, h$  \\
		$f_0$ (s$^{-1}$)            & $1.032 \times 10^{-4}$   & $C_{\rm{o}}$ (W\,m$^{-2}$)             & 310 \\
		$n = 2 L_y / L_x$           & $1.5$               & $T_{\rm{o}}^0$ (K)                     & $285$ \\
		$R_{\rm{E}}$ (km)           & $6370$                   & $\gamma_{\rm{a}}$ (J\,m$^{-2}$ \, K$^{-1}$) & $1.0 \times 10^7$ \\
		$\phi_0$                    & $\pi/4$            & $C_{\rm{a}}$ (W \, m$^{-2}$)             & $C_{\rm{o}}/4$ \\
		$g^\prime$                  & $3.1 \times 10^{-2}$     & $\epsilon_{\rm{a}}$                    & $0.76$ \\
		$r$ (s$^{-1}$)              & $1.0 \times 10^{-7}$     & $\beta$ (m$^{-1}$ \, s$^{-1}$)           & $1.62 \times 10^{-11}$ \\
		$H$ (m)                     & $100$                  & $T_{\rm{a}}^0$ (K)                     & $270$ \\
		$d$ (s$^{-1}$)              & $C/(\rho_o H)$     & $\lambda$ (W\,m$^{-2}$ \, K$^{-1}$)      &   $1004 \, C$ \\
		$k_d$ (s$^{-1}$)            & $(g C)/(\Delta p)$    & $R$ (J\,kg$^{-1}$\,K$^{-1}$)           & $287$ \\
		$k_d^\prime $ (s$^{-1}$)    & $(g C)/(\Delta p)$     & $\sigma$  (J kg$^{-1}$ Pa$^{-2}$)                & $2.16 \times 10^{-6}$ \\   
		$C$ (kg m$^{-2}$ s$^{-1}$)       & $0.008$ & & \\ 
		\hline
	\end{tabular} \label{tab:VDDG}
\end{table}

For the ocean, 

\begin{equation} \label{eq:temp_o}
\gamma_o \left( \frac{\partial T_o}{\partial t} + J(\psi_o, T_o)\right) = -\lambda (T_o-T_a) + E_R,
\end{equation}

with

\begin{equation} \label{eq:rad_o}
E_R = -\sigma_B T_o^4 + \epsilon_a \sigma_B T_a^4 + R_o.
\end{equation}
Here $R_o$ is the shortwave radiation entering the ocean,  $\gamma_o$ the heat capacity of the ocean, and $\lambda$ is the inverse of the time scale associated with heat transfer between the ocean and the atmosphere, which includes both the latent and sensible heat fluxes. In fact, we assume that this combined heat transfer is proportional to the temperature difference between the atmosphere and the ocean. 

The temperatures in both modules are linearized around a reference value in order to reduce the quartic terms of the energy balance equations \eqref{eq:rad_a} and \eqref{eq:rad_o} to linear terms, assuming that the temperature fluctuations are small. This modification helps one to reduce the number of terms on the right-hand side of the ordinary differential equations \eqref{eq:temp_a} and \eqref{eq:temp_o} when building the spectral low-order model. 

The model fields in both its atmosphere and its ocean are developed in Fourier series and truncated at a low order.  The number of modes herein is fixed at 10 for the atmosphere and 8 for the ocean, leading to 20 ordinary differential equations for the former and 16 for the latter. This model configuration is the original VDDG one; see also \cite{Vannitsem2017}. The parameter values used in the present work are listed in Table~\ref{tab:VDDG}. In the configuration of Table ~\ref{tab:VDDG}, there is no substantial low-frequency variability emerging in the extratropical component (See \cite{Vannitsem2022}), allowing for solely clarifying the impact of external low-frequency variability.

\subsection{Modeling the tropical--extratropical interaction} \label{ssec:interact}

\cite{Schemmetal2018} investigated the changes of extratropical wintertime cyclogenesis when El Ni\~no or La Ni\~na events are occurring in the tropical Pacific. They showed, in particular, that the background zonal-flow anomaly is more intense over the north Atlantic during La Ni\~na, while it is stronger over the north Pacific during El Ni\~no. This finding tells us that an important effect of the tropical forcing is to change the intensity of the zonal flow in either region, and that the impact of El Ni\~no and La Ni\~na differs from one region to the other. In order to mimic this dynamic effect in the extratropical VDDG model used here, we impose a direct linear forcing of the model's first barotropic atmospheric mode. It is this barotropic streamfunction mode that represents the intensity of the zonal flow within the atmosphere. Its dynamics is written as

\begin{equation}
\frac{d\psi_{a,1}}{dt} = f_1(\psi_{a,1}, \theta_{a,1}) + g \delta (x+y).
\end{equation} 

Here $f_1(\psi_{a,1}, \theta_{a,1})$ is the original right-hand side of the dynamical evolution \eqref{VDDG-1} of $\psi_{a,1}$; $\delta(x+y)$ represents the eastern tropical Pacific basin's temperature anomalies; and $g$ scales the intensity of the tropical forcing.

Thus $g$ represents the crucial forcing of  the midlatitude VDDG model described in Sec.~\ref{ssec:VDDG} above by the ENSO module of Sec.~\ref{ssec:ENSO}. In our setting, given a positive $g$-value, a positive, warm anomaly will induce an increase of $\psi_{a,1}$,  and hence of the mean zonal flow, $U=- \psi_{a,1} \partial (\sqrt{2} \cos y) /\partial y$. 
This situation corresponds to the intensification of the zonal flow over the north Pacific during an El Ni\~no. If, to the contrary, $g$ is negative, this would correspond to an intensification during La Ni\~na that mimics the ENSO effect over the north Atlantic. In the current analysis, we are focusing on one specific positive value of $g$ providing a strong teleconnection between the tropics and the midlatitudes.

\begin{figure}
\centering
\includegraphics[width=140mm]{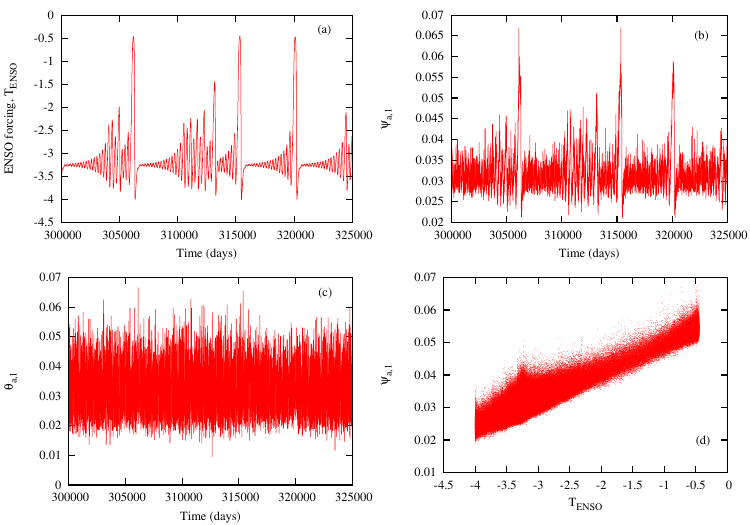}

\caption{Solutions of the coupled tropical-extratropical coupled ocean-atmosphere model for the coupling parameter $g=0.002$. (a) The tropical temperature forcing, $T_{ENSO}$; (b) the forced midlatitude variable, $\psi_{a,1}$; (c) the first baroclinic mode, $\theta_{a,1}$; and (d) the scatter plot of the variables $T_{ENSO}$ and $\psi_{a,1}$.}

\label{solutions}
\end{figure}

Figure \ref{solutions} displays a typical trajectory of the idealized ENSO model for a coupling parameter of $g=0.002$. Panel (a) shows the forcing trajectory of sea surface
temperature in the east equatorial pacific with a succession of strong and less strong El-Niño, and la Niña events, occurring in an irregular manner. The evolution of
the first mode of the barotropic streamfunction, $\psi_{a,1}$, is shown in panel (b), indicating regular bursting associated with strong El-Niño events.
Panel (c) shows the same type of evolution for the first mode of the baroclinic streamfunction, $\theta_{a,1}$, strongly related to the evolution of $\psi_{a,1}$. In
this evolution, the bursting is not visible anymore, indicating a much lower influence of the tropical forcing than for  $\psi_{a,1}$. Finally in panel (d),
the relationship between the instantaneous values of $\psi_{a,1}$ and the temperature in the east part of the tropical ocean basin is displayed, indicating a
strong teleconnection between the modelled tropical Pacific and the streamflow of the midlatitude model. The correlation between the two observables is equal to 0.914, a
strong teleconnection between the two regions in this idealized modelling. If the tropical-extratropical coupling parameter $g$ is modified, one can either increase or
decrease the strength of these teleconnections. Such an impact will be addressed in the next section. 

\begin{figure}
\centering
\includegraphics[width=140mm]{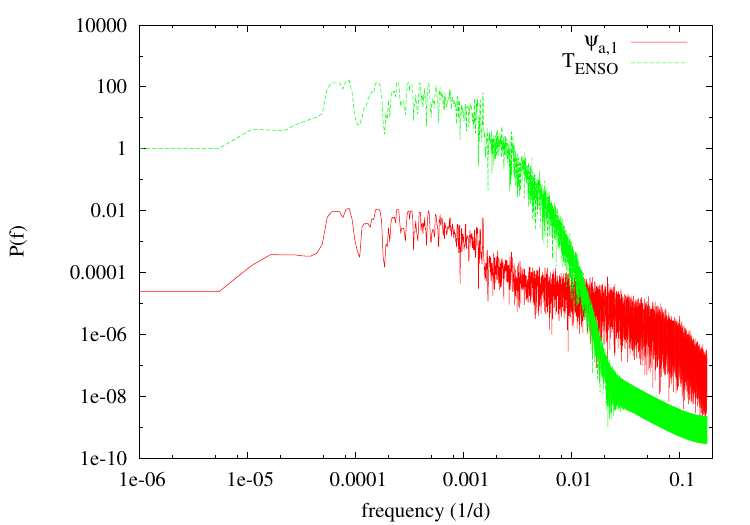}

\caption{Power spectra of $\psi_{a,1}$ and  $T_{ENSO}$ for the coupling parameter $g=0.002$, as obtained using the Multi-Taper Method \cite{Ghil.SSA.2002} with 50000 data points sampled every 2.804 days.}

\label{power}
\end{figure}

In order to further clarify the link between the tropical and extratropical systems, power spectra of $\psi_{a,1}$ and  $T_{ENSO}$ for the coupling parameter $g=0.002$ are displayed in Fig. \ref{power}. For the forcing $T_{ENSO}$, the energy is concentrated at frequencies below 0.001, with a rapid drop at larger frequencies. For $\psi_{a,1}$, the spectrum is much flatter at high frequencies than for the forcing, indicating that a substantial amount of energy is contained at that frequencies. Interestingly a strong similarity of the spectra between frequencies of $10^{-5}$ and $0.002$ is found, indicating the strong link between the forcing and $\psi_{a,1}$. Note that the broadband peak of the $T_{ENSO}$ is obviously reflecting the chaotic nature of the dynamics, but we can isolate a few dominating peaks at about 30, 15, 11 and 5 years. The latter is the one familiar in many time series analysis of the ENSO evolution. The former ones may be emulating the long term behavior of ENSO which displays stronger El-Ni\~no events on decadal and multi-decadal time scales \cite{Wittenberg2009,Guck2017}. 

\section{Predictability at midlatitudes}
\label{sec:predictability}

\subsection{Experimental setup}

The tropical-extratropical coupled system presented in section \ref{sec:equate} allows for a clean analysis of the limits of predictability of the extratropical climate system forced by the tropics. As briefly mentioned in the previous section, the solution generated by the tropical model is chaotic, implying that the signal  to be extracted also exhibits sensitivity to initial conditions as for the extratropics. The Lyapunov time scale of the tropical solution is of the order of 7 years, indicating a very slow increase of errors in time compatible with the dynamics of ENSO. 

A very long control run of the model is first generated which provides the reference against which all forecasting experiments will be evaluated. This constitutes our truth. For the forecasting experiments, it is assumed that an estimate of the truth has been done at specific times along the reference trajectory with some uncertainty. This
uncertainty is generated using random perturbations. We suppose that this uncertainty is coming from a uniform distribution
for the variables in both the tropical region and at midlatitudes. It is also further assumed that we know this distribution exactly and one can therefore generate
an appropriate ensemble forecasting system based on the same distribution as the one used to generate the uncertainty of the observations. The ensemble forecasts are therefore done as follows.

A first set of random perturbations mimicking the uncertainty on the initial conditions in the extratropics is introduced. The distribution used is a uniform distribution of mean 0 and range $[-5. 10^{-7}: 5. 10^{-7}]$. This amplitude is very small as compared
with the amplitude of the variables (in non-dimensional units). Typically of the order of 0.001 $\%$ of the variability of $\psi_{a,1}$. This mimics a situation for which the initial conditions are very well known in the extratropics, that will allow for having a clear picture of the different phases of the error dynamics and the skill of the model variables.

The tropical dynamics is also affected by initial condition errors, that will in turn affect the forecasts in
the extratropics. In order to mimic this feature, a random initial error is also introduced in the initial conditions
of the tropical model, generated from a uniform distribution of zero mean and an amplitude, $PT$, of either 0., 0.02 and 0.05, 0.10, 0.15 and 0.20. Thus the distribution is covering the domain $[-PT/2:PT/2]$. The
perturbation amplitudes are therefore ranging from 0 to 15 $\%$ of the natural variability of the tropical model temperature, $x+y$, used to force midlatitudes.

As in many long-term forecasting systems, an ensemble of forecasts is generated starting from different initial conditions. The number, $M$, of such ensemble members will be modified in order to clarify their impact on the ensemble mean
estimate. The initial conditions of these ensemble members are perturbed consistently with the uncertainties affecting the two models in order to generate a reliable forecast: Once an initial state is defined with the uncertainties mentioned above, random perturbations sampled from the same uniform distributions are generated and added to this initial state to produce each member of the ensemble.

For verification, a few usual metrics are computed: the correlation between the reference trajectory and the forecasts or the ensemble mean (referred as the correlation skill), and the mean square error. To get estimates that are independent of the initial conditions on the attractor
of the coupled system, N=2,000 realizations are performed starting at different location along the reference trajectory.

\subsection{Perfect Teleconnection Predictability}

In order to illustrate the potential impact of the teleconnections introduced in this system on the skill of the solutions, let us first consider the experiments done
without error in the initial conditions of the tropical forcing for values of the coupling parameter $g=0$ and $0.002$. In other words, it is assumed that the forcing evolution is perfectly known, which should allow
for exploiting optimally the teleconnections revealed in Fig. \ref{solutions}. The error in the initial states of the extratropical model is fixed as discussed in the previous section.

\begin{figure}
\centering
\includegraphics{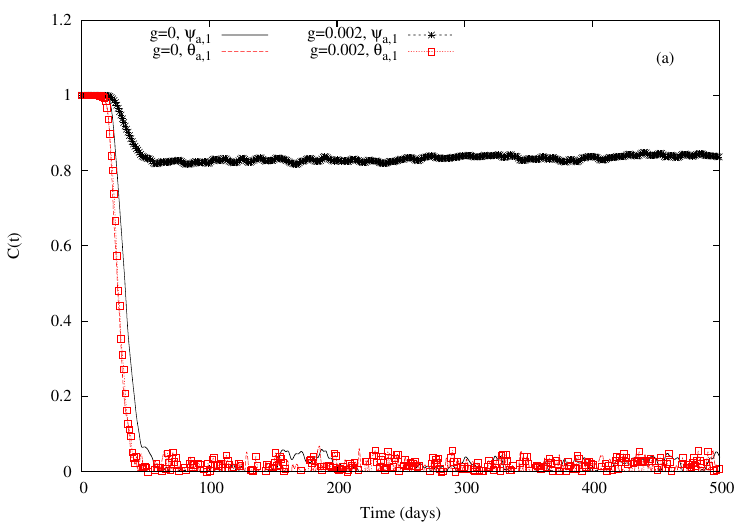}
\includegraphics{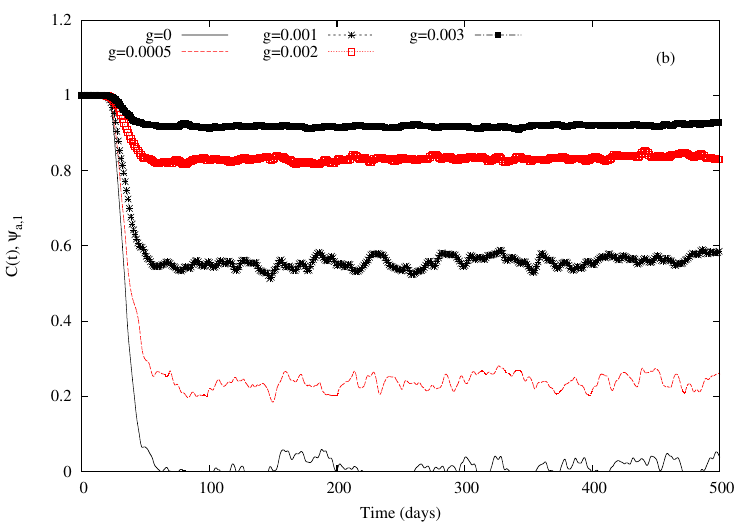}
\caption{Correlation between the reference integration, viewed as perfect observations, and the ensemble mean of forecasts for (a) variables $\psi_{1}$ and $\theta_{1}$ for $g=0.$ and $g=0.002$ and (b) variable $\psi_{1}$ for $g=0.$, $0.001$, $0.002$ and $0.003$. The ensemble size is fixed to $M=1$, hence corresponding to single deterministic forecasts.}
\label{refcorrel}
\end{figure}

Figure \ref{refcorrel}a shows the correlation skill between the control integration (without initial condition errors) and single forecasts ($M=1$) for
variables $\psi_{a,1}$ and $\theta_{a,1}$ for $g=0.$ and $g=0.002$. For $g=0.$, the correlation skill is completely lost after about 50 days. When the forcing
is introduced with $g=0.002$, the picture is similar for the variable $\theta_{a,1}$, while for variable $\psi_{a,1}$ the correlation skill is first quickly
decreasing and then saturates around a value of about 0.82. Two
general comments are in order here with the specific design of the tropical-extratropical model: (i) The teleconnection present between the forcing and $\psi_{a,1}$
induce a long term correlation skill of this variable with a plateau at a level defined by the strength of the teleconnection induced by the tropical forcing; and (ii) there is no obvious propagation of the teleconnection dependence to the baroclinic variable,  $\theta_{a,1}$, strongly related to the dynamics of $\psi_{a,1}$. Although this type
of consideration is not entirely new, it provides a key reference for the experiments that will be done next. In particular, the strong correlation between the
reference and forecasting trajectories for $\psi_{a,1}$ reveals the strong potential of teleconnections. This correlation skill level for $g=0.002$ of $\psi_{a,1}$ will be referred in the following as the {\it Perfect Teleconnection Predictability} (PTP).

Figure \ref{refcorrel}b shows the PTP for various values of $g$ from $0.$ to $0.003$. It reveals that once $g$ is increased, the PTP increases accordingly. It is however interesting to point to the fact that the level of correlation skill shown from, say, $100$ to $500$ days, is always smaller than the correlation between $\psi_{a,1}$ and $T_{ENSO}$: for $g=0.0005$, the correlation is $0.50$, while the level of the corresponding correlation skill in Fig. \ref{refcorrel}b is lower than $0.25$. This reflects the complex interplay between the forcing and the forced variable. So the teleconnection as defined by the correlation between $\psi_{a,1}$ and $T_{ENSO}$ is an upper bound of the actual correlation skill of $\psi_{a,1}$. 

From Fig. \ref{refcorrel}b, it is also interesting to note that whatever the value of $g$, the quick initial decrease of skill toward the plateau is very similar. The reason is that the short term predictability of the extratropical model is almost the same when measured by the first Lyapunov exponent for this range of values of $g$ (see \cite{Vannitsem2022}).  

\subsection{Predictability for M=1}

Let us now focus on the case $g=0.002$ and consider the more realistic situation for which there are some uncertainties in the initial conditions of the tropical forcing, but still with only one ensemble member $M=1$. As the tropical model is
chaotic, it also displays sensitivity to initial conditions that will affect in turn the extratropical forecasts. 

\begin{figure}
\centering
{\includegraphics[width=140mm]{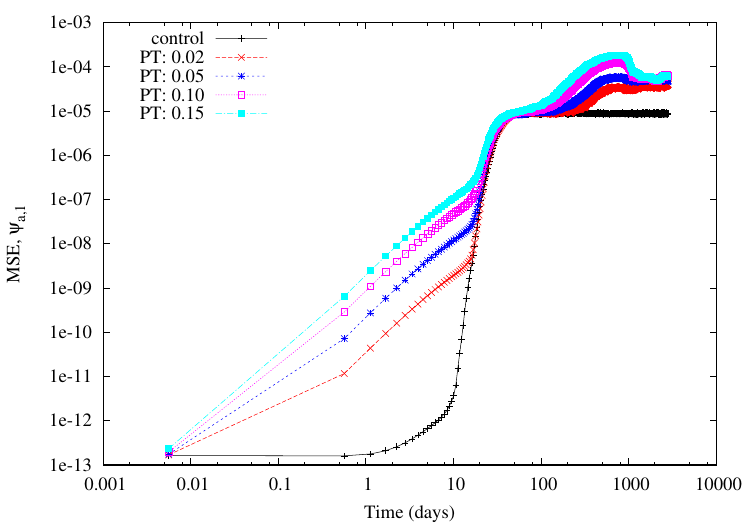}}
\caption{MSE evolution for the variable $\psi_{a,1}$ under the influence of tropical forcing, affected by different initial condition error amplitudes, $PT=0, 0.02, 0.05, 0.10, 0.15$.}
\label{mse-m1}
\end{figure}

Figure \ref{mse-m1} displays the evolution of the Mean Square Error (MSE) for several amplitudes of the perturbations of the initial states of the tropical model. On this doubly logarithmic scale, a first interesting information is the quadratic increase of the MSE for short lead times. This behavior is well documented and typical when model errors or boundary forcing errors are present \cite{Vannitsem2002, Nicolis2003, Nicolis2007}. Interestingly this error quickly amplifies suggesting that boundary forcing errors can considerably harm the forecasts in the extratropics whatever the quality of the initial conditions of the extratropical model. Another important information, is the level of saturation of the error which is reached when forcing errors are present: The error at long lead times is higher. Finally, the way it saturates from, say, 30 days to the end of the forecasts shows a complex feature related to the time scales of the forcing: A plateau is present between 30 to 100 days which is a period for which  the difference between the forcing acting on the reference and the forecasting model is still small, and hence following the error evolution in the absence of any initial condition error of the tropical model; and then followed by an oscillating behavior associated with the typical oscillations found in the original tropical model dynamics. The specific sharp decrease of the error around 900 days needs further clarifications by investigating the detailed evolution of the distribution of the errors as a function of time on these time scales and beyond for various initial error amplitude in the forcing. This question will be addressed in the future. 

\begin{figure}
\centering
{\includegraphics[width=140mm]{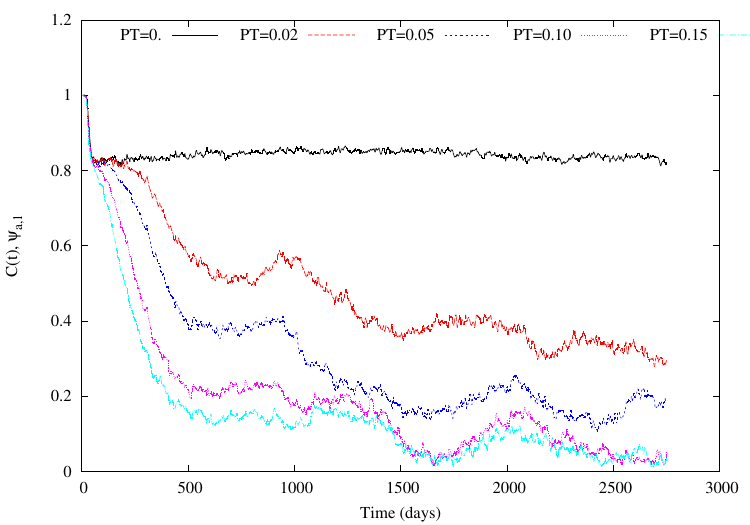}}
\caption{Evolution of the correlation skill for $M=1$ for different amplitudes of the errors introduced in the initial conditions of the tropical model.}
\label{correl-m1}
\end{figure}

Although the error evolution is an interesting quantity to analyze in particular for the short terms, the correlation skill is probably more relevant for the long term in order to clarify the impact of the teleconnections. In figure \ref{correl-m1}, the evolution of the correlation skill between the reference and the forecast of the variable $\psi_{a,1}$ with or without errors in the initial conditions of the tropical model, is displayed. After a quick decrease to the level of the PTP experiment (black continuous line), the correlation skill persists around that level for a few days or weeks depending on the amplitude of the error in the initial conditions of the tropical forcing. It then drops to a lower level that also depends on the amplitude of the initial error. Two important messages here: (i) The teleconnection skill can never be reached on the long term basis when error in the initial conditions are present in the tropical model forcing; (ii) the ability to keep skill on long lead times depends on the amplitude of the errors in the initial conditions of the tropical model.

\subsection{Impact of ensemble averaging}

A first approach to improve the skill is to generate an ensemble of runs starting from different initial conditions and to extract the ensemble mean. It has been shown that the skill is improved even if a very small number of ensemble members is used \cite{Houtekamer1995, Ehrendorfer1997}.

\begin{figure}
\centering
{\includegraphics[width=140mm]{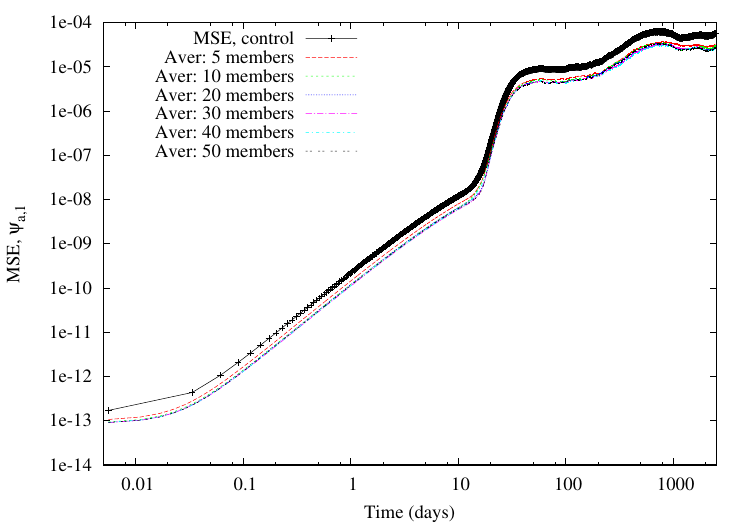}}
\caption{MSE evolution of the forecast of $\psi_{a,1}$ for the control run with $M=1$ (black continuous curve referred as control), and of the ensemble mean for different values of the ensemble size from 5 to 50.}
\label{error-mvar}
\end{figure}

Let us start by investigating the error evolution for different values of M for the case for which the initial condition error amplitude of the tropical model is fixed to $PT=0.05$. The results are displayed in figure (\ref{error-mvar}). The key information that can be extracted is that even with 5-10 members, the main improvement that can be obtained through averaging is reached: little improvements can be obtained going beyond that order. This was already pointed out in \cite{Houtekamer1995} in an intermediate complexity model with much more degrees of freedom. 

\begin{figure}
\centering
{\includegraphics[width=140mm]{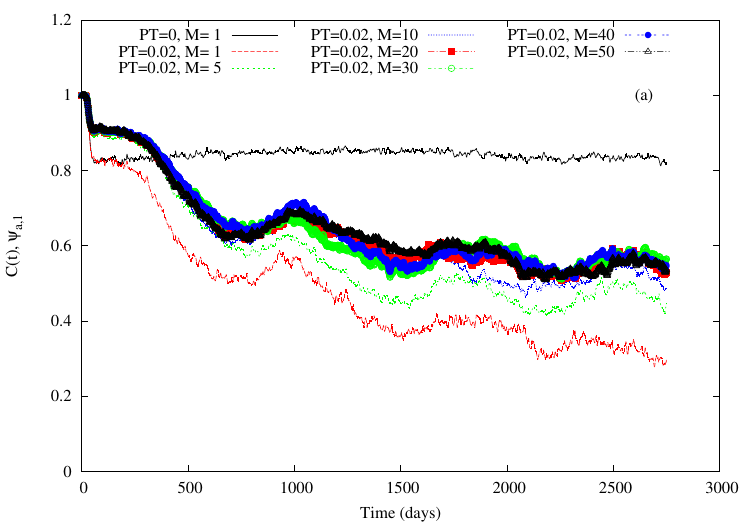}}
{\includegraphics[width=140mm]{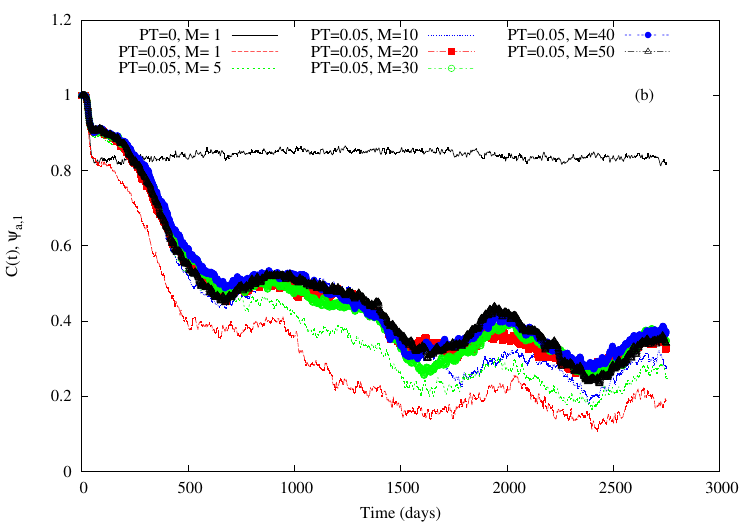}}

\caption{Correlation skill of the forecasts for (a) $\psi_{a,1}$ with a amplitude perturbation of $PT=0.02$, (b) $\psi_{a,1}$ with a amplitude perturbation of $PT=0.05$. Within each panel, the curves are representing different number of members M in the estimation of the ensemble mean. For reference, the case without errors in the initial conditions of the tropical model for $M=1$ is also displayed (black continuous curve).}
\label{correlSCI-psia}
\end{figure}

This impact of ensemble averaging on the correlation skill is then illustrated in figure
\ref{correlSCI-psia} for (a) $\psi_{a,1}$ with a perturbation amplitude of $PT=0.02$, and (b) with a perturbation amplitude of $PT=0.05$. In these plots,
the correlation skill corresponding to the PTP is also displayed, as well as the case with $M=1$. 

A first remark is that the correlation skill between the reference and the ensemble mean forecasts is still lower than PTP for long lead times when initial condition perturbations are introduced in the forecasting model of the forcing. Moreover when the number of members is increased, the quality of the forecast is increased, but already 10 members are enough to reach the optimal skill. The latter result is consistent with the fact that the ensemble mean is rapidly converging to its asymptotic value as discussed in \cite{Houtekamer1995}. 

A second important result is the improvement of the correlation skill when the ensemble mean is used as forecast between 50 and about 300 days. The correlation skill is now reaching values close to the teleconnection correlation between the forcing and the variable. In other words, the ensemble averaging is properly filtering the fast weather variability around the mean. The ensemble mean is here an appropriate approach to isolate the signal present in the variable $\psi_{a,1}$.   

\begin{figure}
\centering
{\includegraphics[width=140mm]{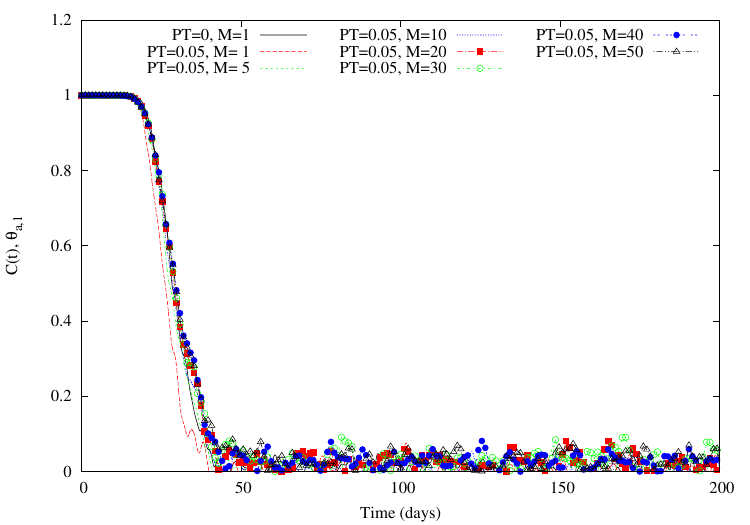}}
\caption{As in figure \ref{correlSCI-psia} but for the variable $\theta_{a,1}$ with an amplitude perturbation of $PT=0.05$. The curves are representing different number of members $M$ in the estimation of the ensemble mean. For reference the case without errors in the initial conditions of the tropical model is also displayed (black continuous curve).}
\label{correlSCI-theta}
\end{figure}

For the variable $\theta_{a,1}$, the skill decreases quickly and is similar whatever the initial condition error in the tropical model as illustrated in figure \ref{correlSCI-theta}, suggesting the weak sensitivity of this variable to the external forcing. Note however that there is some improvements by taking ensemble averages in the phase of quick decrease of the skill around 30-40 days, even with a small number of ensemble members.

In the last decades, many studies have tried to evaluate the potential forecasting skill of climate forecasting models, and compared it to the actual forecast skill \cite{Kumar2014, Imada2021, Yang2023}. The potential forecasting skill is the correlation between the ensemble mean and any member of the ensemble. In this context if the model is perfect and if the model is properly initialized, the potential predictability should be equal to the actual predictability. 

\begin{figure}
    \centering
    \includegraphics[width=140mm]{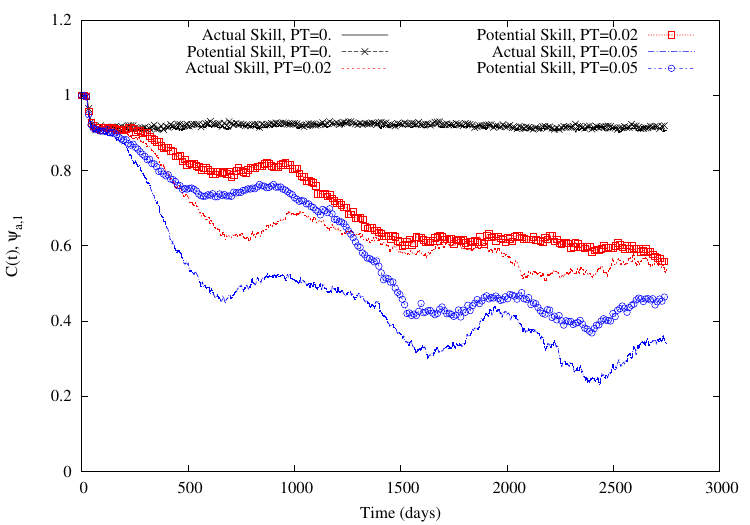}
    \caption{Comparison of the actual forecasting skill of the ensemble mean and the potential predictability -- as defined by the skill of the ensemble mean in forecasting any of the member of the ensemble. Three cases are displayed with or without error in the initial conditions of the tropical model.}
    \label{potpredict}
\end{figure}

In fact even if the model is properly representing the reference evolution, a discrepancy between the actual skill and the potential skill is found in the current experiment. This is illustrated in figure \ref{potpredict}. The potential skill is larger than the actual skill, except when there is no initial condition errors in the tropical model forecasts (black curves). This feature in the current coupled tropical-extratropical model may only be understood by the fact that even if perturbing randomly and symmetrically the initial conditions around the observations (and hence the reference) in the tropical model, a forecasting bias is emerging (not shown). This forecasting bias is most probably due to the way the solutions of the forecasting model are converging back to the model's attractor. Such feature is very subtle and is worth addressing further in a future studies by analyzing the temporal properties of these transient trajectories, and to propose new approaches for perturbing the model more consistent with the natural dynamics of the reference system as discussed for instance in details in \cite{Demaeyer2022} or using analog initialization as in \cite{Li2013}.  

\subsection{Impact of temporal averaging}

Ensemble averages are often used to improve the estimate the the future state of the system. As shown in the previous section, this is indeed interesting and the correlation skill score can be improved to a certain extent. But once the number of members is large enough, no improvement is obtained, this number being related to the quick convergence of the ensemble mean to its asymptotic value. 

Another way to extract the signal from the forecasts is often to take temporal averages (running means). This question has been addressed in several papers in the past, in particular showing that the short term dynamics of the error as described by the dominant Lyapunov exponent is the same as for instantaneous fields \cite{Nicolis1995, Vannitsem1997, Vannitsem1998}. The main modification that can then expect is for long term forecasts as discussed in \cite{Roads1987}, but the gains are very low when investigating atmosphere-only models \cite{Tribbia1988}. We also explored this question in the context of the tropical-extratropical coupled model to clarify the impact on the time averaging in the extraction of the signal associated with the tropical forcing. 

\begin{figure}
    \centering
    \includegraphics[width=140mm]{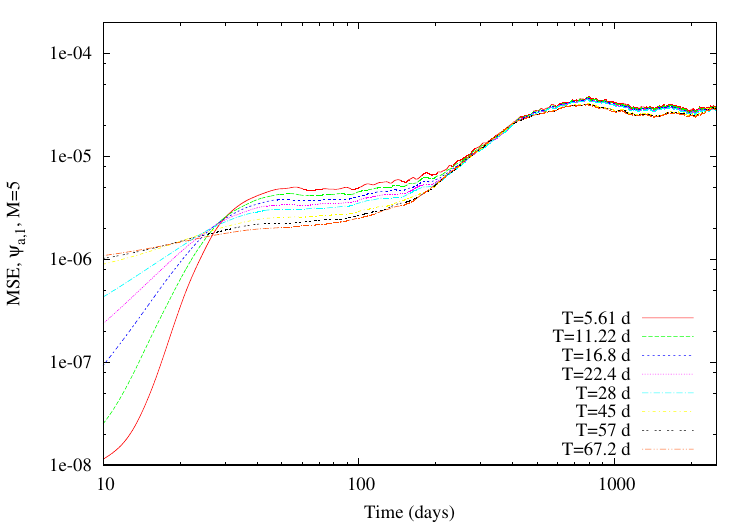}
    \caption{MSE evolution for $\psi_{a,1}$ with $M=5$ and different temporal averaging from $T=5.61$ d to $T=67.2$ d. Note that the lead time Time$=0$ corresponds to the end of the averaging period of the forecast.}
    \label{error-timeav}
\end{figure}

\begin{figure}
    \centering
    \includegraphics[width=140mm]{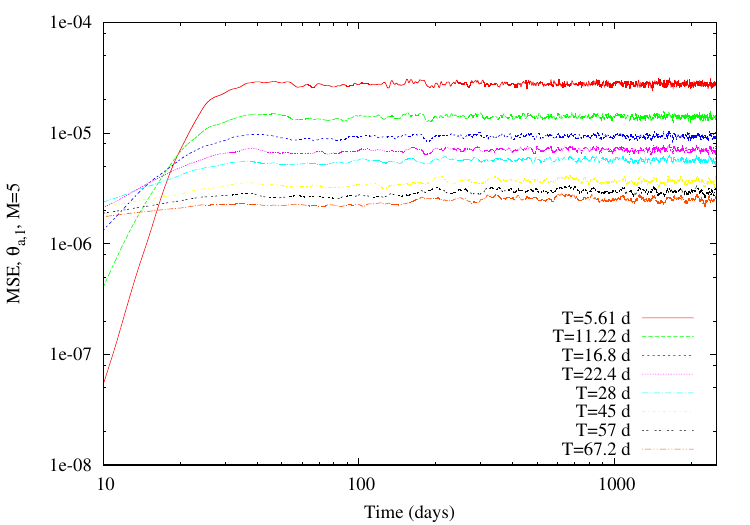}
    \caption{MSE evolution for $\theta_{a,1}$ with $M=5$ and different temporal averaging from $T=5.61$ d to $T=67.2$ d. Note that the lead time $=0$ corresponds to the end of the averaging period of the forecast.}
    \label{error-timeav-theta}
\end{figure}

Figure \ref{error-timeav} is displaying the different MSE curves for the ensemble mean with $M=5$ for the temporal averages of the variable $\psi_{a,1}$. The averages have been performed from about 5 days to 60 days. Note that the lead time, Time$=0$ days, is taken at the end of the averaging period $T$, so that Time$=0$ on the figure is associated with the averaging over $T$ days of the original forecast. Four different phases can be found in the evolution of the error: first an overall increase of the MSE for short times associated with the averaging of successive situations displaying large errors as compared with the reference for longer lead times; a second phase during which the MSE amplitude is decreasing when $T$ becomes large (from lead times of 25 to 200 days); a third phase of rapid growth (from 200 days to 400 days) for which averaging does not change the MSE amplitude; and finally a phase (after 400 days) for which the MSE amplitude again decreases for large values of $T$. This complicated dependence reflects that time averaging may indeed reduce the variance, and hence the error, of the processes but not uniformly for all the time scales. This aspect was already amply discussed in \cite{Nicolis1995}, and shows up here in different responses of the MSE phases to averaging. This differential behavior as a function of the phase does not however shows up for $\theta_{a,1}$ as illustrated in figure \ref{error-timeav-theta} with a constant decrease of the MSE saturation level as a function of time averaging. As $\theta_{a,1}$ is much less affected by the forcing than $\psi_{a,1}$, this differential behavior for $\psi_{a,1}$ could be associated with the action of the tropical forcing. 

\begin{figure}
    \centering
    \includegraphics[width=140mm]{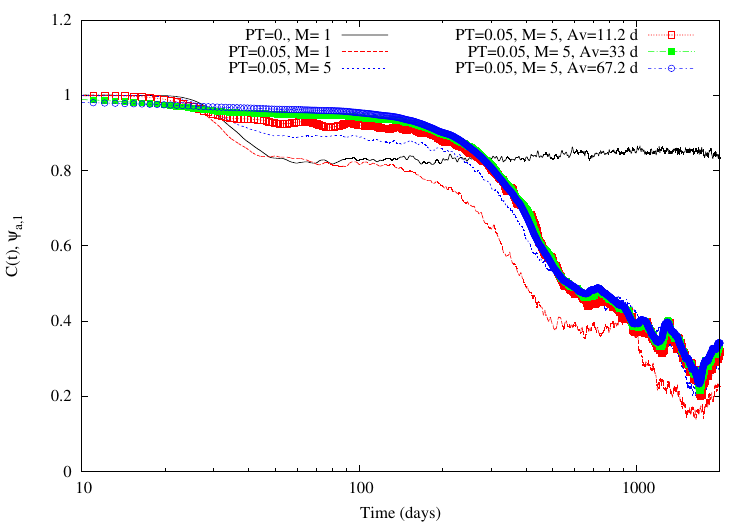}
    \caption{Evolution of the corrrelation skill for $\psi_{a,1}$ for various combinations of averaging (ensemble and/or in time).}
    \label{correl-timeav}
\end{figure}

\begin{figure}
    \centering
    \includegraphics[width=140mm]{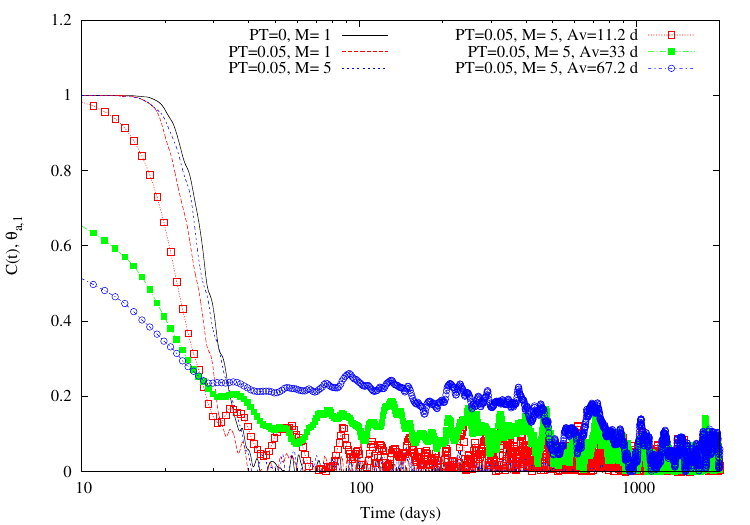}
    \caption{Evolution of the corrrelation skill for $\theta_{a,1}$ for various combinations of averaging (ensemble and/or in time).}
    \label{correl-timeav-theta}
\end{figure}

Another very useful view of what is happening while taking temporal averaging is to evaluate the changes in the correlation skill. In figure \ref{correl-timeav}, the evolution of the correlation skill for the ensemble mean of 5 members ($M=5$) of $\psi_{a,1}$ without time averaging, and with three different time averages of 11.2, 33 and 67.2 days. A small gain of averaging is found between 50 and 300-day lead times, suggesting the usefulness of this averaging at such lead times. If now one looks at the corresponding figure for $\theta_{a,1}$ (Fig. \ref{correl-timeav-theta}), the picture is very different. Although the instantaneous quantities, with or without ensemble averaging, do not show skill beyond 40 days, the temporal average allows for getting skill of this observable at long lead times: For the temporal averaging of 67.2 days, a correlation skill of the order of 0.2 is found up to almost 1000 days. This suggests that the temporal averaging is appropriately removing the high frequency variability of $\theta_{a,1}$ and isolate the signal associated with the influence of the tropical forcing.   

To sum up, temporal averages can be very helpful in extracting the skill of observables influenced by external forcing signals, even when the teleconnections are weak.

\section{Conclusions}

Low-frequency variability coming from the presence of an external slowly-evolving forcing are often isolated through teleconnections. In the present work, we have investigated in an idealized context the impact of the presence of teleconnections between the tropics and the extratropics on the limits of predictability of the extratropical fields. This has been done in a coupled tropical-extratropical model that was recently built to evaluate the emergence of LFV in the extratropics and the properties of the corresponding pullback attractors \cite{Vannitsem2022}. 

A highly idealized twin experiment was performed by running the coupled tropical-extratropical model with initial condition errors for both components of the system, and comparing its forecasts to a long reference run. A few generic properties were found:

\begin{itemize}
    \item If there is no error in the initial conditions of the tropical model, the forced variable, here the first barotropic atmospheric mode, shows a skill on very long time scales thanks to the slowly-varying external forcing (which is in this case the same for the model forecast and the reference). 
    \item When errors in the initial conditions are present, the correlation skill is considerably degraded at interannual lead times. The degradation of skill also depends on the amplitude of the initial condition errors in the tropics, suggesting that accurate initial conditions are necessary to get the full potential of teleconnections.
    \item Ensemble averaging may improve the quality of the forecasts, but the number of members necessary to reach an appropriate skill does not need to be large: 5-10 members are enough in this perfect model framework.
    \item Temporal averaging can also improve the quality of the forecasts (at the expense of course of being able to forecast extreme events), in particular when the forcing affects weakly the observable under interest. In the current context, the first baroclinic mode shows a very weak teleconnection (correlation) with the tropical forcing, and when averaging, a strong correlation skill is emerging.     
\end{itemize}

This work also showed that even if the model is perfect and the initial observation distribution is known, the actual correlation skill of a forecast is always less or equal than the potential correlation skill as defined using the forecasting model itself only. This feature is very subtle and is most probably associated with the way the model forecasts are converging back to the system's attractor. This attractor is the same as the reference, but the path to converge to it is generating a forecasting bias which reduces the actual correlation skill. This suggests that appropriate initial conditions of the tropical model generating the forcing should be looked for, allowing for reducing this bias. This will be the subject of a future work along the lines of \cite{Demaeyer2022}.

In the current work, the low-frequency variability is introduced through a one way external forcing from the tropics to the midlatitudes. Things become more complicated when part of the low-frequency variability is generated intrinsically in the forced system as for instance in \cite{Pierini2020} or when the forcing is interacting in a complicated way with the intrinsic low-frequency variability like in \cite{Vannitsem2022}. In both cases, the mean may loose significance, and a full probability distribution is then necessary to characterize the dynamics of the reference system. Further investigations of such complex interactions are worth performing in the future.

Finally this problem should also be tackled in the context of a more realistic model for which the initial conditions in the tropical domain should be modified in order to clarify how well we can exploit the teleconnections for prediction purposes. This will also be the subject of a future work. 

\ack
This work is supported by the Belgian Policy Office under Grant B2/20E/P1/ROADMAP, a contribution to the ROADMAP project of the JPI-Climate-JPI-Ocean research activities.
%%%%%%%%%%%%%%%%%%%%%%%%%%%%%%%%%%%%%%%%%%
%%%%%%%%%%%%%%%%%%%%%%%%%%%%%%%%%%%%%%%%%%
\bigskip

\bibliographystyle{unsrt}

\bibliography{teleconnections-new}

\end{document}